\documentclass[sigchi-a,nonacm]{acmart}

\AtBeginDocument{%
  \providecommand\BibTeX{{%
    \normalfont B\kern-0.5em{\scshape i\kern-0.25em b}\kern-0.8em\TeX}}}

\setcopyright{acmcopyright}
\copyrightyear{2018}
\acmYear{2018}
\acmDOI{10.1145/1122445.1122456}

\acmConference[Woodstock '18]{Woodstock '18: ACM Symposium on Neural
  Gaze Detection}{June 03--05, 2018}{Woodstock, NY}
\acmBooktitle{Woodstock '18: ACM Symposium on Neural Gaze Detection,
  June 03--05, 2018, Woodstock, NY}
\acmPrice{15.00}
\acmISBN{978-1-4503-XXXX-X/18/06}

\usepackage{paralist}
\usepackage{fancyhdr}
\usepackage[absolute]{textpos}
\begin{document}
\title{Extremism \& Whataboutism: A Case Study on Bangalore Riots}

\author{Saloni Dash}
\affiliation{
\institution{Microsoft Research}
\city{Bengaluru}
\country{India}
}
\author{Gazal Shekhawat}
\affiliation{
\institution{London School of Economics and Political Science}
\city{London}
\country{United Kingdom}
}
\author{Syeda Zainab Akbar}
\affiliation{
\institution{Microsoft Research}
\city{Bengaluru}
\country{India}
}
\author{Joyojeet Pal}
\affiliation{
\institution{Microsoft Research}
\city{Bengaluru}
\country{India}
}

\begin{abstract}
A common diversionary tactic used to deflect attention from contested issues is whataboutery which, when used by majoritarian groups to justify their behaviour against marginalised communities, can quickly devolve into extremism.
We explore the manifestations of extreme speech in the Indian context, through a case study of violent protests and policing in the city of Bangalore, provoked by a derogatory Facebook post. Analyses of the dominant narratives on Twitter surrounding the incident reveal that, most of them employ whataboutism to deflect attention from the triggering post and serve as breeding grounds for religion-based extreme speech. We conclude by discussing how our study proposes an alternative lens of viewing extremism in the Global South. 
\end{abstract}



\maketitle

\begin{textblock*}{\paperwidth}(2.54cm,8in)
      CSCW'21 Extremism Research Workshop
\end{textblock*}

\section{Introduction \& Background}
The increased use of social media has had notable effects in democratising communication, with individuals being able to connect with people from all over the world as well as their leaders and representatives. Yet, the same stride raises concerns about the polarised networks these platforms can promote, through media manipulation, influence and misinformation. As our work on the use of social media in Indian politics \cite{dash2021divided, mishra2021rihanna} has found, users' experience of social media, and the ``public opinion" that flows through it is mediated, and a few influential accounts can dominate the discourse surrounding a political event. We attempt to study extremism in India through the analysis of prominent narratives propagated by influential actors during a political crisis.

In order to study extremism in the Indian context, one must take into account the fragmented nature of Indian society, fractured on the basis of religion, caste and linguistics. The incident we analyse, speaks of mob violence from the Muslim community in the southern parts of Bangalore, DJ Halli. The riots were triggered by the nephew of a legislative member of the Indian National Congress, who allegedly created a derogatory post of Prophet Muhammad on Facebook. Police action to control the protests led to firing on the crowd, where three people died. Along with questions about religion and policing, the fact that the legislative member belonged to the Dalit community also adds the intersection of caste to the riot.

The Bangalore riots case is well suited to contextualise extreme speech studies in the Indian setting.
We find that a disproportionate amount of literature on extreme speech focuses on the USA or Western Europe. Moreover, a threat to western democracies is seen as a starting point of interest in non-western regions. Therefore, most studies on extreme speech in the Global South are viewed in the context of the ``default" American cyberspace instead of understanding them on their regionalised terms. Through our study of social media narrative on the Bangalore Riots, we aim to posit conceptual questions that multi-ethnic regions such as India can ask of extremism research. 

Specifically, we study how the concept of ``to quoque" or whataboutery, a common strategy used to misdirect attention from a controversy and shift the debate to accusations and counter-accusations, breeds extremism in the Indian Twitter sphere. Fundamentally, the employment of a counter-instance against an opponent, draws charges of hypocricy against them, as they are said to be inconsistent in their concerns \cite{bowell2020whataboutisms}. Thus, whataboutism can lead an audience to dismiss the speaker's initial argument, on grounds such an inconsistency, while the speaker is forced to defend on a newer plane of arguments \cite{bowell2020whataboutisms}. As \citet{little2017politics} have noted, the urgency to draw parallels to grievances between presumably opposing sides, can lead to a false equivalence of injustices different communities have faced. 
In India, whataboutery is a common sight in TV debates, where producers strive to choreograph objectivity by pitting guests against each-other, as strategic noise-making \cite{mishra2018broadcast}.

\section{Methodology}
\begin{margintable}
\centering
\caption{Distribution of Tweets by Hashtag}
\label{tweets_dist}
\begin{tabular}{ll}
\toprule
\textbf{Hashtag}    & \textbf{\# of Tweets} \\
\midrule
\#bangaloreriots    & 157,699               \\
\#bangaloreviolence & 51,086                \\
\#bengaluruburns    & 18,791                \\
\#DJHalliViolence   & 2264     \\
\#bengalururiots    & 61,475                \\
\bottomrule
\end{tabular}
\end{margintable}
\subsection{Data}
We collect data for the case of Bangalore Riots, that occured on the night of 11th August 2020, from Twitter. We collect tweets using the Twitter search API by using the top 5 trending hashtags related to Bangalore Riots as keywords. We collect a total of 291k tweets for the period of 11th August, 2020 to 18th August, 2020. The hashtags and the corresponding number of tweets collected are given in Table \ref{tweets_dist}. Since tweets contain links to other sites, emoticons, non-alphanumeric characters etc., we pre-process the tweets in the following manner --
\begin{compactenum}
    \item Removal of hyperlinks \& emojis.
    \item Removal of retweet (RT @) and user mentions (@)
    \item Removal of punctuations and non-alphanumeric characters.
    \item Case fold all characters to lower case.
\end{compactenum}



\subsection{Network Analysis}
We construct a directed retweet graph from the collected Twitter data with all the users who tweeted about it as nodes, $ v \in V$. A directed edge $e_{ij}$, with weight $w_{ij}$ exists from user $v_i$ to user $v_j$ ($v_i, v_j \in V$) if $v_i$ has retweeted $v_j$ $w_{ij}$ times ($w_{ij} \geq 1$). The retweet graph is then visualised by using the Force Atlas 2 layout in Gephi\cite{gephi2009}, which sets gravity rules for node attraction, using their edge weights to make the graph easier to interpret and visualise.  

\subsection{Topic Modelling}
In order to extract the dominant narratives related to the riots on Twitter, we use Latent Dirichlet allocation (LDA), an unsupervised topic modelling algorithm \cite{blei2003latent}. LDA is a generative probabilistic model of a corpus, where documents are represented as random mixtures over latent topics and each topic is characterized by a distribution over words. While LDA has been shown to be highly successful with long documents, it does not perform well with shorter documents which are sparse and have very limited information about co-occurring words \cite{wang2006topics}. A common approach used to address this has been to combine short texts to build larger pseudo documents based on user data or bookmarking features like hashtags \cite{mehrotra2013improving}.

For the purpose of our study, we combine all tweets related to the event into one document. We train a LDA model on this corpus where the optimal number of topics is determined by a grid search on the values between 5 and 50. We pick the model with the highest coherence value \cite{roder2015exploring}. We then manually assess each topic returned by the model, by studying the keywords and accompanying tweets for the topic. We use our contextual knowledge of the event to pick topics that are most relevant to the riots.

\section{Findings}
\subsection{Key Players}
\begin{marginfigure}
  \centering
  \includegraphics[width=\linewidth]{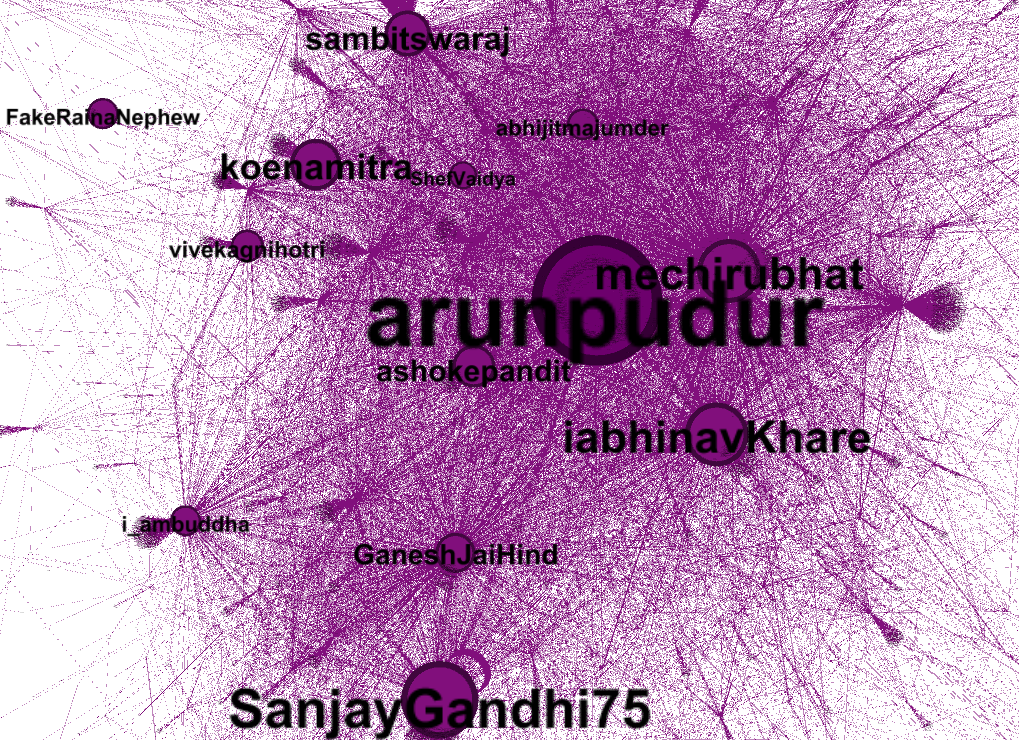}
  \caption{Retweet Network}
  \label{rt_network}
\end{marginfigure}

As described previously, we plot the retweet network constructed from the tweets in Figure \ref{rt_network}. The size of the nodes is proportional to their degree , i.e. the bigger the node, the more central is its position in the retweet network.
\newline \indent
The account with the highest degree centrality is @arunpudur, a buisnessman with over 73k followers on Twitter. One of his top retweeted tweets include ``\textit{How many people were attacked and cities burnt when an entire movie degrading Hindu Gods was done by Muslim Amir Khan? Does this justify mob lynching and riots? \#bangaloreriots \#Bangalore}". 
This is a distinguishing characteristic of whataboutery, where one deflects criticisms of their group's actions by drawing comparisons to the other group's actions and establishing superiority over them. In this case, the derogatory post on Prophet Muhammad, which is said to have triggered the Bangalore Riots is being defended by arguing that in a similar instance, when Hindus were provoked by Muslims, they did not react to it by attacking people and damaging public property.

We see that several of the top retweeted accounts engage in a similar rhetoric. @koenamitra's tweet ``\textit{...we tolerate but called ‘intolerant’..they kill but they’re the most peaceful right?...JNU product had called Ma Durga a prostitute. Sexy Radha toh suna hoga?....}" or @AsYouNotWish's tweet ``\textit{\#BangaloreRiots are a very disproportionate response to a deleted post on Prophet Muhammad. Hindu Gods are made fun of day and night, not only by ordinary Muslims but also by their leaders like Owaisi. We don’t start rioting....}".
While the tweets themselves do not qualify as extreme speech, they trigger extreme speech against Muslims. In response to such tweets, we see tweets referring to Muslims as ``hooligans", ``barbarians"  and ``terrorists".
We discuss other similar narratives and other central accounts like @sambitswaraj, @ashokepandit etc. whose tweets directed the discourse on the riots, in the section below.


\subsection{Dominant Narratives}

\begin{marginfigure}
  \centering
  \includegraphics[width=\linewidth]{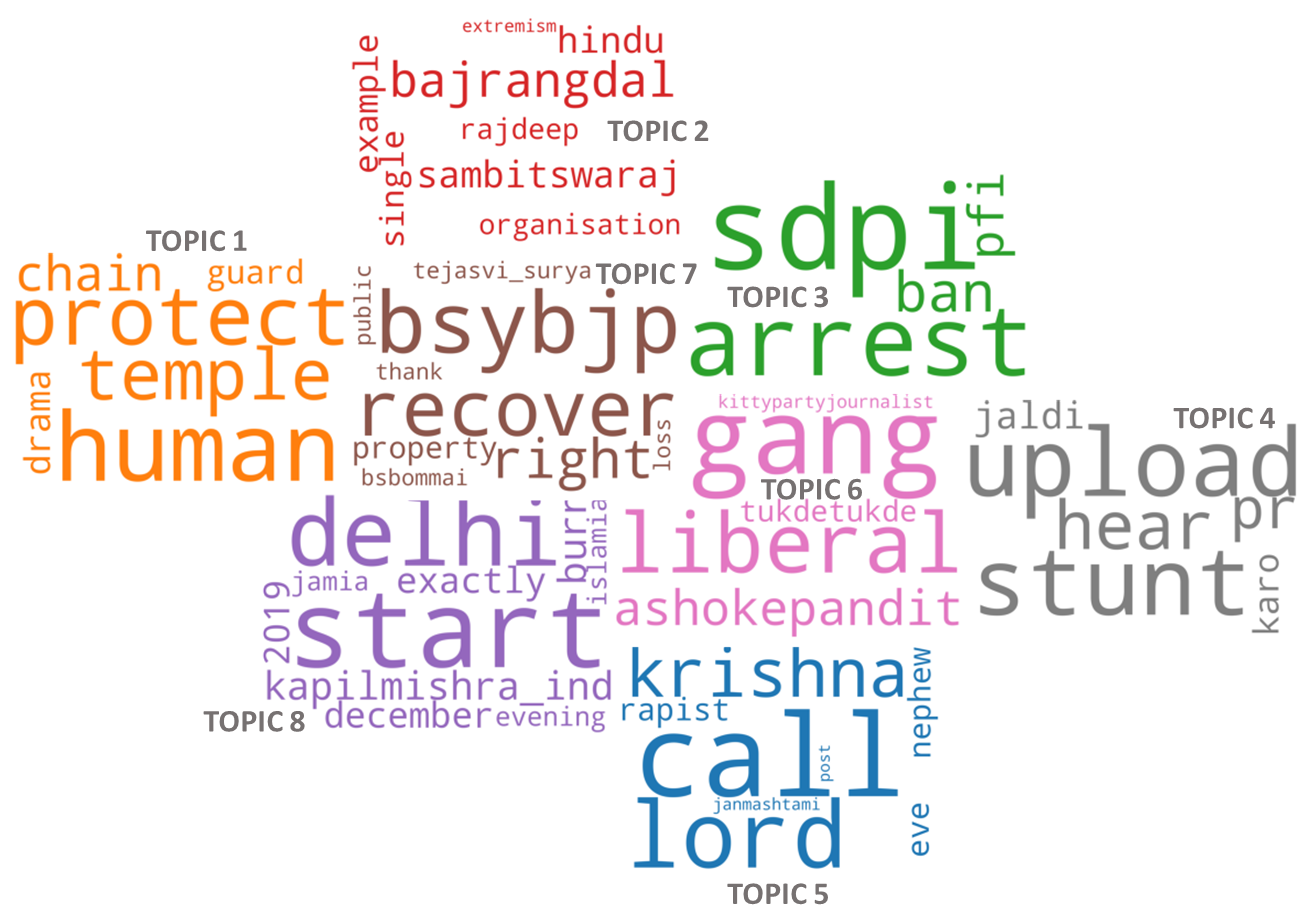}
  \caption{Wordclouds of Topics}
\end{marginfigure}

\begin{compactitem}
    \item \textbf{Human Chain (Topic 1):} A human chain was formed around a Hindu temple by some members of the Muslim community in order to protect it from escalating riots and show solidarity with the Hindu community. While the gesture was applauded on Twitter by many, others viewed it as ironic and hypocritical, with tweets using insect imagery like ``\textit{How cute! Ants protecting the sugar grains....}" or ``\textit{..human chain to protect chicken from non vegetarians}".
    
    \item \textbf{Bajrangdal Comparison (Topic 2):} This narrative was triggered by @sambitswaraj's tweet ``\textit{bajrangdal...show me a single example of any “Hindu” organisation burning police stations \& trying to kill police officers..}". The riots in this case are used to defend a Hindu nationalist organization by arguing that they never resorted to the type of violence that the rioters engaged in, thereby establishing the increased tolerance levels of Hindus as compared to Muslims.
     \item \textbf{SDPI Involvement (Topic 3):} A theory that was widely discussed on Twitter was the involvement of the Social Democratic Party of India (SDPI). It was posited that the riots were not spontaneous, but orchestrated by leaders of SDPI, who were dissatisfied by the government's policies on issues regarding Jammu \& Kashmir, Citizenship Amendment Act etc. 
     Tweets urged government officials to ban SDPI, touting it as a front for terrorist organizations. This narrative is again used to trivialise the initial post that triggered the riot, by claiming that the riots were a coordinated attempt at deepening existing communal tensions.
    \item \textbf{PR Stunt (Topic 4):} A widely discussed theory, stemming from the human chain incident was that the people who formed the chain to protect the temple only did it as a PR stunt. The grounds for this theory was mainly a video that was uploaded by those guarding the temple, at the end of which somebody allegedly said ``\textit{..video upload karo jaldi}" or upload the video fast. The tweets in which these these theories were discussed, ``\textit{\#pissful community ke log...clearly hear someone saying :- \#video \#upload \#kar \#jaldi}", were punctuated by terming Muslims as ``pissfulls", a play on ``peacefuls" which is used  in a derogatory manner in this context.
    \item \textbf{Retaliation (Topic 5):} The widely accepted belief was that the derogatory post on Prophet Muhammad was in retaliation to a Muslim's post defaming a Hindu deity. A viral tweet by @FakeRainaNephew (suspended since then), with over 4k retweets claimed ``\textit{A Muslim had shared a post calling Lord Krishna a rapist on the eve of Janmashtami. Angry with that this nephew of Congress MLA posted that Prophet was rapist...}". 
    These discussions triggered intense hate speech against the Muslim community, labelling them as ``terrorists" and ``rapists" with the hashtag \#ProphetWasRapist becoming viral in this context.
    \item \textbf{Liberals' Silence (Topic 6):} The tweet by @ashokepandit  ``\textit{Will the pseudo liberals, \#intolerance gang, \#TukdeTukde gang,  \#KhanMarket gang and \#KittyPartyJournalists condemn \#Bengaluru riots? \#BangaloreRiots}", spurred criticisms of the silence of the politicians and journalists who usually defended the Muslim community.
    \item \textbf{Riot Damages (Topic 7):} The state government's decision to recover damages to public property from the accused rioters was widely celebrated on Twitter with parliamentarian @TejasviSurya lauding the move ``\textit{... deciding to recover all losses to public property by attaching the private property of rioters...}". This can be better understood in the context of the anti-CAA protests, where Yogi Aditynath, Chief Minister (CM) of Uttar Pradesh seized assets of the rioters in order to recover the losses to public property. Hailed as the ``Yogi Model", tweets that displayed their support for this move branded the rioters as ``venom", ``fanatic dogs" and ``jihadis".
    \item \textbf{Links to Other Riots (Topic 8):} Parallels were drawn to other riots, including the Jamia Millia Islamia anti-CAA protests on 13th December, 2019 with @KapilMishra\_IND tweeting and gaining over 8k retweets ``\textit{This is Bangluru last evening...Exactly How They started burning Delhi in December 2019 from Jamia Islamia 
\#bangaloreriots...}. This should also be viewed in the context of Delhi riots, because the author of this tweet, Kapil Mishra was accused of inciting the Delhi riots in which Muslims were disproportionately killed. However, the incident of Banaglore riots was used to trivialise his actions by his supporters, who claimed that he was just a ``scapegoat" for the Delhi riots. Moreover, as indicated by the tweet, the Bangalore riots are seen as part of a larger conspiracy in which Muslims coordinate attacks and riots against Hindus in order to spread communal hatred and violence.
    
\end{compactitem}





\section{Discussion}
Our findings indicate that the discourse on communal violence in India is teeming with whataboutery, where the majority's grievances in this instance were used to justify the triggering post and negate organized violence against Muslims in the past. As literature on the phenomenon notes, whataboutery risks turning societal tensions into zero-sum conflicts, where a crescendo of antagonisms replaces the need for resolutions. 
A preliminary analysis of prime time TV debates surronding the riots also resemble whataboutery on Twitter, where partisan spokespersons are quick to resort to extreme speech. 
The mediascape around Bangalore riots evidences that dehumanizing speech is not just the tool of fringe groups, which are usually framed at odds with the ``liberal" democracies informing mainstream extremism research. As multi-ethnic societies such as India attest, progression into extremism is not linear as the ``radicalisation" hypothesis would suggest, but the entrenchment of hateful narratives instead reflects the social dysfunction upon which social media is built and utilized. Thus, our study presses upon us to reflect, how do we grapple with a society where extremism is no longer extreme?

\bibliographystyle{ACM-Reference-Format}
\bibliography{ref.bib}
\end{document}